\documentclass[preview, 12pt, 3p, sort&compress]{elsarticle}

\usepackage{amssymb}
\usepackage{amsmath}
\usepackage{multirow}
\usepackage{booktabs}
\usepackage{todonotes}
\usepackage{subcaption}
\usepackage{hyperref}

\usepackage{xcolor}
\usepackage{amsmath}
\usepackage[normalem]{ulem}

\journal{Materials \& Design}

\begin{document}

\begin{frontmatter}

\title{Segregation to grain boundaries in disordered systems: an application to a Ni-based multi-component alloy}

\author[inst1]{Dominik Gehringer}
\author[inst1]{Lorenz Romaner}
\author[inst1]{David Holec}

\affiliation[inst1]{organization={Department of Materials Science, Montanuniversität Leoben},
            addressline={Franz-Josef-Stra{\ss}e 18}, 
            city={Leoben},
            postcode={8700}, 
            country={Austria}}

\begin{abstract}
Segregation to defects, in particular to grain boundaries (GBs), is an unavoidable phenomenon leading to changed material behavior over time.
With the increase of available computational power, unbiased quantum-mechanical predictions of segregation energies, which feed classical thermodynamics models of segregation (e.g., McLean isotherm), become available.
In recent years, huge progress towards predictions closely resembling experimental observations was made by considering the statistical nature of the segregation process due to competing segregation sites at a single GB and/or many different types of co-existing GBs.
In the present work, we further expand this field by explicitly showing how compositional disorder, present in real engineering alloys (e.g. steels or Ni-based superalloys), gives rise to a spectrum of segregation energies.
With the example of a $\Sigma 5$ GB in a Ni-based model alloy (Ni-Co-Cr-Ti-Al), we show that the segregation energies of Fe, Mn, W, Nb, and Zr are significantly different from those predicted for pure elemental Ni.
We further use the predicted segregation energy spectra in a statistical evaluation of GB enrichment, which allows for extracting segregation enthalpy and segregation entropy terms related to the chemical complexity in multi-component alloys.
\end{abstract}

\begin{graphicalabstract}
\includegraphics[width=12.5cm, height=5cm]{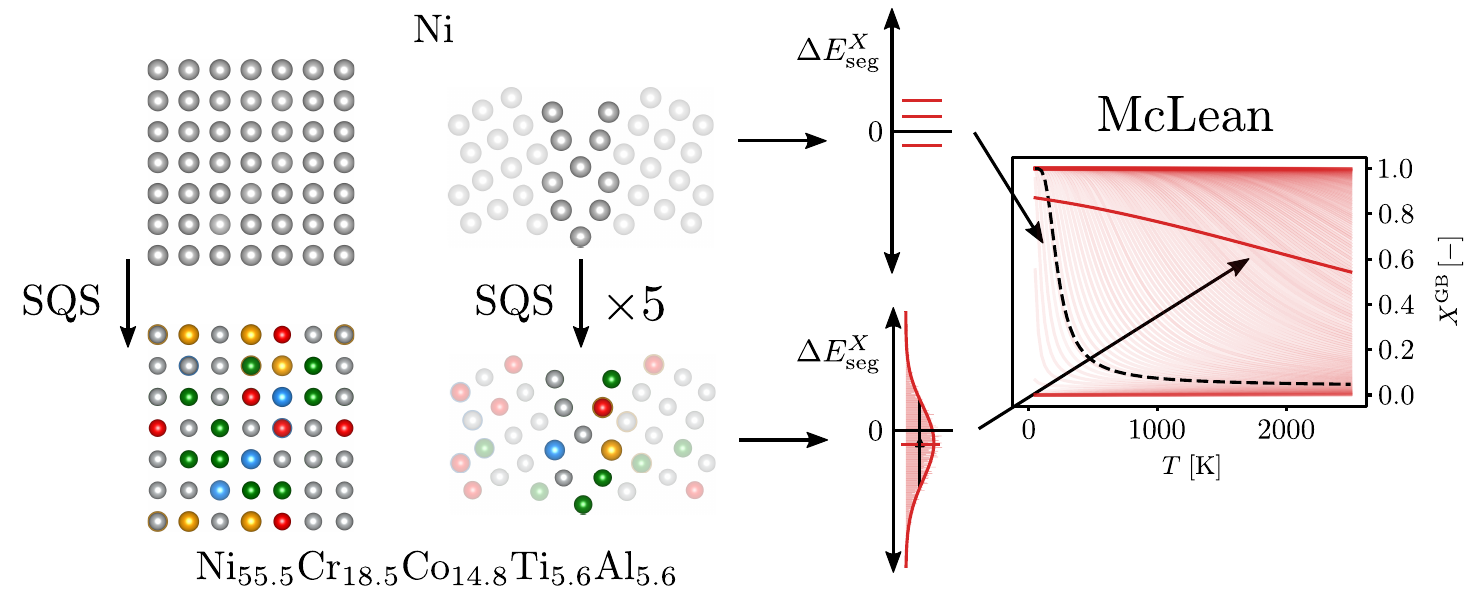}
\end{graphicalabstract}

\begin{highlights}
    \item Study of solute segregation behavior of Fe, Mn, W, Nb and Zr to a $\Sigma 5 (210)[001]$ GB in a Ni-based model alloy
    \item The use of disordered atomistic models gives rise to the segregation energy spectrum
    \item Chemical complexity of multi-component alloy leads to stronger segregation behavior than pure Ni
\end{highlights}


\end{frontmatter}


\section{Introduction}
\label{sec:sample1}

Grain boundary (GB) segregation is a key phenomenon that needs to be understood and controlled when developing novel materials. This issue has been addressed experimentally for many decades. 
With ever-growing computational power, quantum-mechanical methods have become routinely used in the last two decades to predict GB segregation energies. 
Nevertheless, \emph{ab initio} methods, such as e.g. density functional theory (DFT), remain computationally very demanding. 
Consequently, three major approximations are commonly imposed on the atomistic models. 

Firstly, only high symmetry grain boundaries are treated by DFT due to restrictions on the model size. This limitation can be overcome by applying some special methods such as a combination of quantum mechanics and molecular mechanics~\cite{Huber2016, Huber2017}; however, such techniques are still computationally costly and not yet widely used.
A more common approach is to employ (semi-)empirical interatomic potentials (IP) together with molecular mechanics to investigate segregation to structurally complex GBs~\cite{Huber2018, Aksoy2021} or even simulating polycrystals~\cite{Wagih2019, Wagih2020, Wagih2022, Reichmann2024-xo}.
A recent publication of \citet{Wagih2024-gx} demonstrated that, indeed, the spectral treatment stemming from the structural variety of grain boundary sites in polycrystals is crucial for quantitative modeling of segregation.

Secondly, most studies---as is the case also of the present contribution---restrict themselves to the dilute limit. 
When studying concentration dependence~\cite{Scheiber2018} or co-segregation~\cite{Mai2022, Garrett2022, Sakic2024}, many combinations of GB states need to be sampled, which leads to a drastic increase in the number of calculations involved. 

The third simplification is that segregation phenomena in alloys are always---except for a work by \citet{Scheiber2015} and an even very recent study~\cite{Umashankar2023}---modeled with a system of the corresponding pure metal. 
For example, pure iron is used as a surrogate model to steels for predicting the segregation behavior~\cite{Ito2020, Peng2023, Hu2020, Mai2022, Sakic2024}. 
Similarly, pure Ni is used for \emph{ab initio} modeling of segregation in Ni-based superalloys~\cite{Razumovskiy2011b, Xue2021a, Xue2021b, Ebner2021, Taji2022, Xiao2021}. 

While the spectral nature of segregation energies due to structural variety in realistic materials has been heavily advocated in the recent literature~\cite{Wagih2019, Wagih2020, Wagih2022, Wagih2024-gx, Scheiber2021}, the statistical distribution of segregation energies due to complex-chemistry in the matrix has been touched upon only scarcely and only for binary systems~\cite{Scheiber2015, Umashankar2023}. 
Therefore, in the present study, we make a first step towards addressing segregation of solute species in the dilute limit in a \emph{multi-component matrix}. 
Our aim is to discuss the effect and modeling challenges of chemical disorder caused by a multi-component matrix. 
The composition of the matrix itself is considered to be fixed in the bulk and the GB, and we leave a full treatment of the segregation phenomena of the matrix elements for future studies.

We have chosen a high symmetry $\Sigma5(210)[001]$ GB (Fig.~\ref{fig:model-disordered}) as a structural model for our first-principles study.
This is motivated by the vast existing literature on segregation in Ni to this GB, which helps discuss the effect of the complex chemical composition.
The matrix composition, inspired by realistic Ni-based superalloy compositions~\cite{Choudhury1998-zm}, e.g. the Udimet 720 and Rene 65 alloys~\cite{Strondl2012-iy} or ABD-850AM developed for additive manifacturing~\cite{Tang2023-nk}, is summarized in Tab.~\ref{tab:nominal-composition}. 
Note that this composition is also close to the matrix $\gamma$ phase after the $\gamma'$ phase precipitation in certain cases~\cite{Tang2023-nk}.
We considered five majority (matrix) elements, Ni, Cr, Co, Ti, and Al (whose composition is fixed, i.e., their segregation is not considered), and focused on the segregation behavior of five minority Fe, Mn, W, Nb, and Zr species. 
We model the matrix as a solid solution using special quasirandom structures (SQS)~\cite{Wei1990}. 

\begin{table}[h]
    \centering
    \renewcommand{\arraystretch}{1.2}
    \begin{tabular}{lcccccl}    
        \toprule
        \multirow{2}{*}{Alloy} & \multicolumn{5}{c}{Major (matrix) species (at.~\%)}\\
         & Ni & Cr & Co & Ti & Al & \multirow{2}{*}{Ref.}\\ \midrule
         model composition & 55.6 & 18.5 & 14.8 & 5.6 & 5.6 & this work\\
         Udimet 720 & 54.9 & 17.3 & 14.3 & 5.9 & 5.2 & \cite{Strondl2012-iy}\\
         Rene 65 & 54.9 & 17.9 & 12.8 & 4.5 & 4.5 & \cite{Strondl2012-iy}\\
         ABD-850AM & 51.3 & 21.1 & 17.8 & 3.1 & 3.2 & \cite{Tang2023-nk} \\
         ABD-850AM$^\star$ & 46.5 & 23.3 & 22.1 & 1.6 & 3.2 & \cite{Tang2023-nk} \\\bottomrule
         \multicolumn{6}{l}{$^\star$ composition of the $\gamma$ phase only}
    \end{tabular}
    \caption{Comparison of the model alloy composition of selected Ni-based alloy. Overall composition unless stated otherwise.}
    \label{tab:nominal-composition}
\end{table}

The paper is structured as follows: 
We start with revising the standard methodology for computing the segregation energy in pure elemental systems. 
We continue by introducing the necessary theory for the spectral representation of segregation energy, similar to Refs.~\cite{Huber2018, Wagih2019, Scheiber2021, Aksoy2021, Wagih2022}, but point out the major difference---the origin of the spectra (chemical vs. structural complexity). 
The theory part is concluded by generalizing the McLean model~\cite{McLean1957} for a multi-component system. 
In the next part, we describe our atomistic models and the DFT setup.
The results section starts by comparing segregation phenomena in the pure system with our multi-component setup. 
We continue by comparing the predictions of the McLean model for the two cases. 
The final section is related to extracting enthalpy and entropy of segregation, which have been phenomenologically used before to explain experimental observations~\cite{Lejcek2016}.
We conclude with a summary of the main results.

\section{Theory}
\subsection{Segregation energy in disordered systems}

\subsubsection{Segregation energy in a pure elemental system}
\begin{figure}
    \centering
    \includegraphics[width=0.875\textwidth]{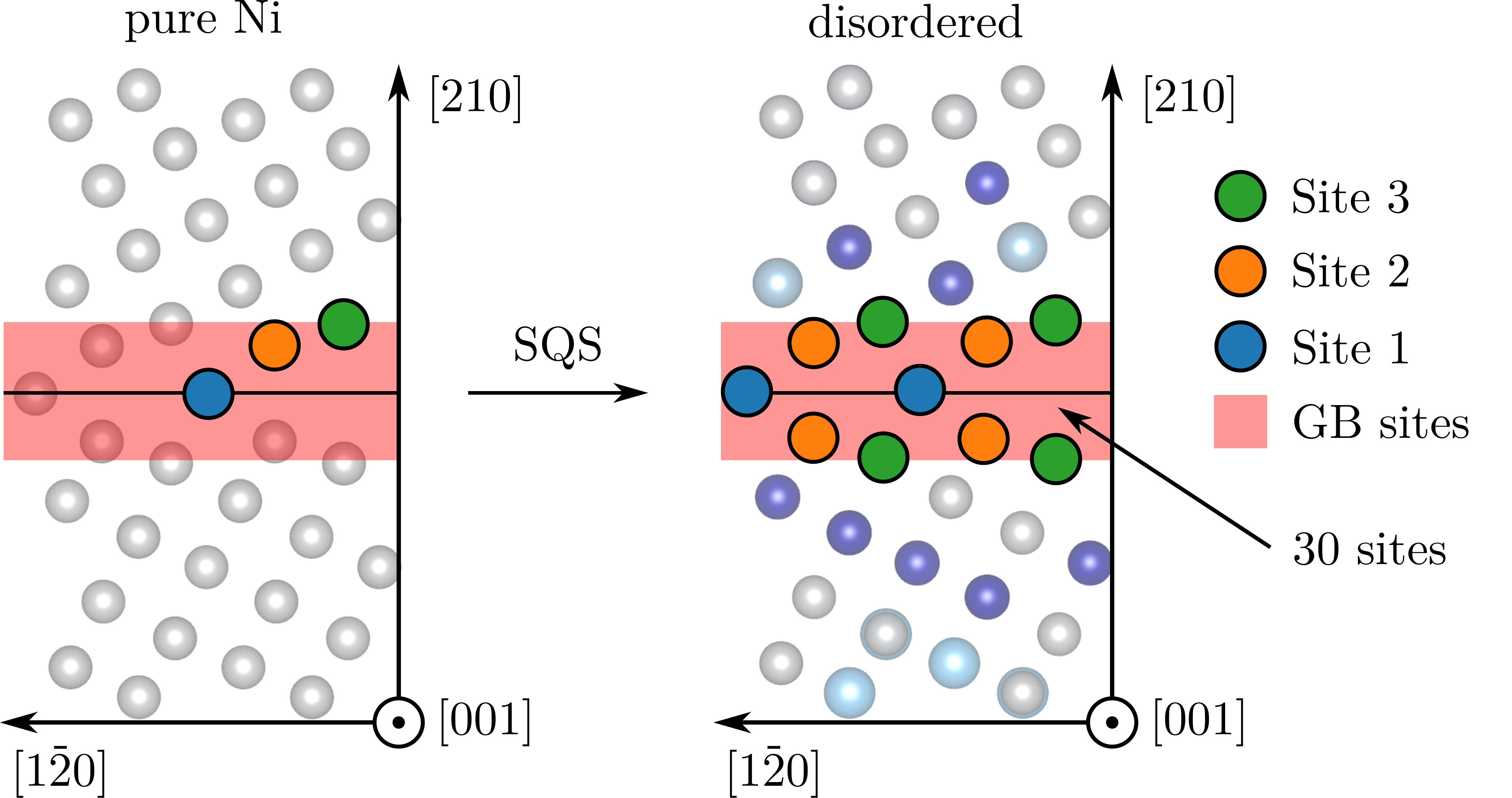}
    \caption{Atomistic model of a pure Ni GB (left) and a disordered GB. 
    The colored circles schematically illustrate the GB sites. 
    The red zone refers to the region that we define as the GB.}
    \label{fig:model-disordered}
\end{figure}

The segregation energy quantifies the thermodynamic driving force for segregation to a GB. 
This is evaluated as a difference between the formation energy of a point defect of the solute species $X$ in bulk (B), $E^{f, X}_{\text{B}}$ with the corresponding formation energy at the grain boundary (GB), $E^{f, X}_{\text{GB}}$. 
For a GB model of pure system of species $M$ where the solute atom $X$ sits at the GB site $i$, the segregation energy reads:
\begin{align}
    \Delta E^{X,i}_{\text{seg}} &= \underbrace{\left(E_{\text{GB}}^{X \rightarrow M_i} - E_{\text{GB}} + \mu^{M} - \mu^X\right)}_{E^{f, X}_{\text{GB}}} - \underbrace{\left(E_{\text{B}}^{X \rightarrow M} - E_{\text{B}} + \mu^{M} - \mu^X\right)}_{E^{f, X}_{\text{B}}} \notag \\
    &= E_{\text{GB}}^{X \rightarrow M_i} - E_{\text{B}}^{X \rightarrow M} - E_{\text{GB}} +  E_{\text{B}}\ .
    \label{eq:segregation-energy-pristine}
\end{align}
Here, $E_{\text{GB}}$ and $E_{\text{B}}$ refer to the total energies of the undecorated systems. 
$\mu^M$ and $\mu^X$ refer to the chemical potentials of matrix and solute species, respectively. 
$E^{X \rightarrow M}_{\text{B}}$ and $E^{X \rightarrow M_i}_{\text{GB}}$ are the energies of the decorated system, where one $M$ atom is replaced by an $X$ atom. 
We note that in a steady state, the chemical potential of each species is the same in bulk and at the GB~\cite{Kwiatkowski_da_Silva2019-sl}. 
It is a common practice to define a set of GB sites (sites which are belonging to the GB).
Index $i$ labels symmetry inequivalent GB sites. 
For example, for the pure Ni $\Sigma 5$ GB in Fig.~\ref{fig:model-disordered} can take $i \in \{1, 2, 3\}$.
All other sites have bulk-like nearest-neighbor environments.
We remind the reader that the segregating species $X$ differ from the matrix elements $M$ in our treatment.
This will allow us to apply the McLean formalism of segregation isotherms (Sec.~\ref{sec:McLean_isotherms}), which is valid in dilute limit; the matrix elements have all concentrations far beyond the dilute limit (cf. Tab.~\ref{tab:nominal-composition}).
Consequently, the composition of matrix species in the bulk and the GB is identical (within what a discrete atomistic model allows (Sec.~\ref{sec:gb-models}).

\subsubsection{Disorder in the bulk: $E^{f,X}_\text{B}$}
In the following two sections, we shed light on the implications of chemical disorder on the meaning of segregated energy. 
Firstly, the matrix of a multi-component system is composed of many species $\widetilde{M} \in \mathbf{M} = \{\text{Ni}, \text{Cr}, \text{Co}, \text{Ti}, \text{Al}\}$, in contrast to a pure Ni matrix. 
Secondly, even for one species $\widetilde{M}$, the bulk formation energy is not a single-valued scalar since the chemical disorder breaks the symmetry and introduces a variety of local environments. 
In Eq.~\eqref{eq:segregation-energy-pristine}, we implicitly considered a single bulk site that we compare with many GB sites. 
In the case of a multi-component matrix, we also get a spectrum of bulk formation energies, $E^{f, X \rightarrow \widetilde M_j}_\text{B}$, denoting a solute $X$ replacing the matrix species $\widetilde M$ at the site $j$:
\begin{equation}
    E^{f, X \rightarrow \widetilde M_j}_\text{B} = E_{\text{B}}^{X \rightarrow \widetilde M_j} - E_{\text{B}} + \mu^{\widetilde M} - \mu^X \ .
\end{equation}

\subsubsection{Disorder at the GB: $E^{f,X}_\text{GB}$}
\label{theory:disorder-at-gb}
Already in the pure elemental case, there are several different GB sites (indexed with $i$, Fig.~\ref{fig:model-disordered} left) differing by the spatial arrangement of their neighboring sites. 
Similarly as for the bulk, the chemical disorder breaks also the symmetry at those GB sites causing that each might have different species as its neighbors.
We note that the GB zone (red region in Fig.~\ref{fig:model-disordered}) containing the GB sites was chosen in accordance with Refs.~\cite{Xue2021b, Razumovskiy2015}. 
As a consequence, the site index for the GB state in the multi-component case may correspond to any of these GB sites (instead of three (symmetry-inequivalent) in the pure elemental case, Fig.~\ref{fig:model-disordered}). 

In short, for both formation energies $E^{f,X}_\text{B}$ and $E^{f,X}_\text{GB}$ are sets of values rather than a single value:
\begin{equation}
     E^{f, X}_{\text{B}/\text{GB}} = \left\{E^{X \rightarrow \widetilde M_i}_{\text{B}/\text{GB}} - E_{\text{B}/\text{GB}} + \mu^{\widetilde M} - \mu^X \right\}_{\forall i \in 1, \ldots, N_{\text{B}/\text{GB}}}
     \label{eq:formation-energy-sets}
\end{equation}
where $N_{\text{B}}$ and $N_{\text{GB}}$ are the number of bulk and GB states respectively. 

Consider now that the solute atom can occupy any bulk state, and from there cam reach each GB state. 
Consequently, we obtain the segregation energy spectrum by creating all possible combinations of sites and elements in sets in Eq.~\eqref{eq:formation-energy-sets}:
\begin{align}
    \Delta E^{X_{ij}^{\widetilde M \widetilde M'}}_{\text{seg}} &=  E_{\text{GB}}^{X \rightarrow \widetilde M_i} - E_{\text{GB}} - E_{\text{B}}^{X \rightarrow \widetilde M_j'} + E_{\text{B}} + \mu^{\widetilde M} - \mu^{\widetilde M'} \notag \\
    &= E_{\text{GB}}^{X \rightarrow \widetilde M_i} - E_{\text{GB}} - E_{\text{B}}^{X \rightarrow \widetilde M_j'} + E_{\text{B}} + \underbrace{\Delta\mu^{\widetilde M \widetilde M'}}_{\mu^{\widetilde M} - \mu^{\widetilde M'}} \ .
    \label{eq:segregation-energy-general}
\end{align}

In order to avoid any ambiguity related to a particular choice of chemical potentials (see later discussion of Fig.~\ref{fig:formation-energies-splitted}), which is a non-trivial task in the case of compositionally complex alloy, we will further treat only situations where $\Delta\mu^{\widetilde M \widetilde M'}=0$, i.e. when $\widetilde M=\widetilde M'$. This also helps to preserve the compositions of our bulk and GB simulation boxes as similar as possible due to their rather small size (100--200 as restricted by the DFT calculations). We also note that in the framework of this work, we do not consider any co-segregation of minority species, nor do we aim to discuss segregation competition between solute (minority) and matrix (majority) elements; segregation of the latter is not considered at this level of simplification.
Thereby we finally arrive at:
\begin{equation}
    \Delta E^X_{\text{seg}} = \left\{ \Delta E^{X_{ij}^{\widetilde M \widetilde M'}}_{\text{seg}} \right| \left.\vphantom{\Delta E^{X_{ij}^{\widetilde M}}_{\text{seg}}} \; \widetilde M = \widetilde M ',\  i=1,\ldots,N_\text{GB},\ j=1,\ldots,N_\text{B}\right\}\ .
    \label{eq:segregation-energy-definition}
\end{equation}
The above Eq.~\eqref{eq:segregation-energy-definition} is a definition for the \emph{segregation energy in a multi-component matrix}.
It considers all possible swaps between bulk and grain boundary sites occupied by the same chemical species.

\subsection{Distribution of segregation energy}
The definition of segregation energies by Eq.~\eqref{eq:segregation-energy-definition} results in a spectrum of values, which we will conveniently represent by a corresponding distribution function. 
Such an approach was suggested already by \citet{White1977} and got attention again more recently. 
For example \citet{Huber2018}, sampled the fundamental zones~\cite{Homer2015} (orientations of the cutting boundary plane~\cite{Patala2011,Patala2013}) of a $\Sigma 5$ grain boundary. 
Therein, they used a Gaussian distribution to describe the energy spectra. 
\citet{Scheiber2021} discussed the impact of energy spectra more intensively and tried to connect it with the measurable enthalpy and entropy. 
They found a Gumbel distribution fitting their energy spectra best. 
Similarly, \citet{Wagih2019} studied the grain boundary segregation in a polycrystal and investigated the impact of the segregation energy spectra on the stability of nanocrystalline materials. 

In the present work we follow the approach of \citet{Wagih2019} and use a skew-normal distribution to represent the segregation energies. 
For a solute $X$, the skew-normal distribution reads
\begin{equation}
    \hat{F}^X(\Delta E_\text{seg}^X) = \dfrac{1}{\sigma \sqrt{2\pi}} \exp\left(-\frac{(\Delta E_\text{seg}^X - \mu)^2}{2\sigma^2}\right)\text{erfc}\left(-\frac{\alpha(\Delta E_\text{seg}^X - \mu)}{\sqrt{2\sigma^2}}\right)
    \label{eq:dis-skew-normal}
\end{equation}
where $\mu$ and $\sigma$ are the parameters of a Gaussian distribution, and $\alpha$ the skewness parameter. 
The sign of $\alpha$ determines the side of the skew; $\alpha = 0$ yields a Gaussian distribution. 
These parameters are obtained by fitting Eq.~\eqref{eq:dis-skew-normal} to a histogram of the (discrete) segregation energies from Eq.~\eqref{eq:segregation-energy-definition}. 

We further define the mean value of the distribution as:
\begin{align}
    \left<\Delta E_\text{seg}^X\right> = \left<\hat{F}^X\right> = \int_{-\infty}^\infty \hat{F}^X(\Delta E_\text{seg}^X) \Delta E_\text{seg}^X \text{d}\Delta E_\text{seg}^X \ .
    \label{eq:dis-expectation-contious}
\end{align}
The width of a spectrum is quantified by its full-width at half maximum (FWHM). 

\begin{figure}[h]
    \centering
    \includegraphics[width=0.85\textwidth]{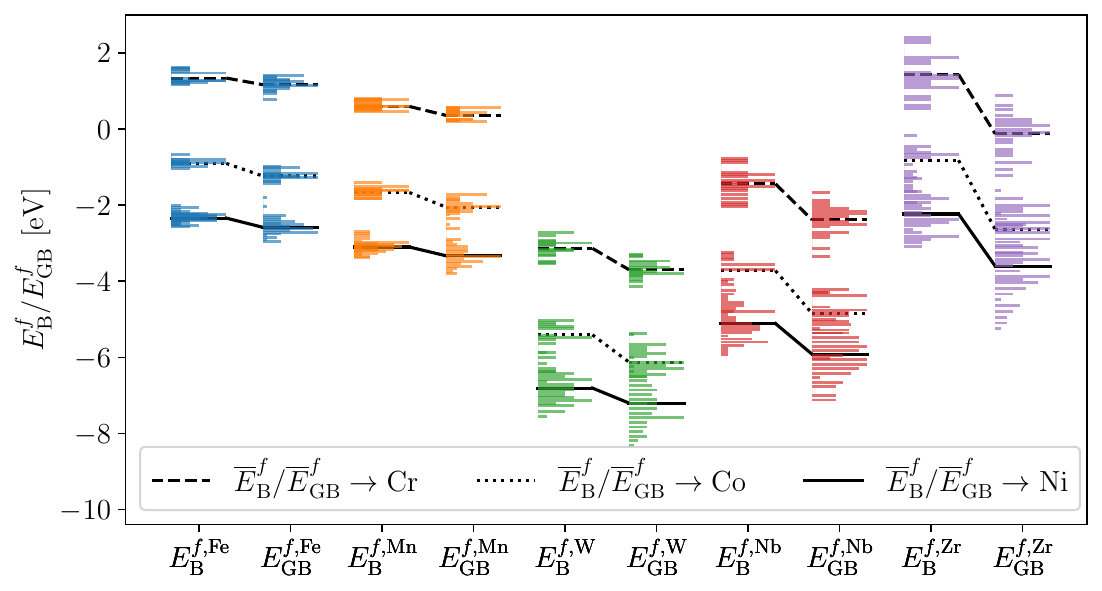}
    \caption{Bulk formation ($E^{f,X}_\text{B}$) as the left and GB formation energy spectra $E^{f,X}_\text{GB}$ as the right column for each solute (shown by different colors). 
    The spectra for each solute replacing Ni (solid),  Co (dotted), and Cr (dashed) sites are plotted individually. 
    For the individual formation energies (Eq.~\eqref{eq:formation-energy-sets}), the chemical potential terms do not cancel out; hence the individual spectra are shifted along the $y$-axis by a constant offset $\mu^X - \mu^{\widetilde M}$ with $\widetilde M \in \{\text{Ni}, \text{Co}, \text{Cr}\}$.}
    \label{fig:formation-energies-splitted}
\end{figure}

We want to point out a qualitative difference between the energy spectra discussed in literature~\cite{Huber2018, Wagih2019, Scheiber2021} and the present study. 
The spectra in previous studies originated from sampling many structurally different GBs in a pure metal. 
In contrast, we focus only on a single GB ($\Sigma 5(210)[001]$), however, in a compositionally disordered system. 
The two main implications are as follows: 
Firstly, the chemical disorder gives rise to a distribution of bulk formation energies, too, compared to a single state in the chemically pure bulk case. 
This is illustrated by Fig.~\ref{fig:formation-energies-splitted}. 
For each solute (different color), Fig.~\ref{fig:formation-energies-splitted} shows a pair of formation energy distributions (exhibiting the same color). 
The bulk states' energy distribution on the left, and the GB states' distribution on the right. 
These distributions further split into subsets depending on the substituted matrix species $\widetilde M\in\{\text{Ni}, \text{Co}, \text{Cr}\}$.
Secondly, the segregation energy distribution arises from chemically different local environments of bulk and GB states. 
In other words, the distributions in literature~\cite{Huber2018, Scheiber2021, Wagih2019, Aksoy2021} are caused by the structural variety of grain boundaries, while here, they are caused by the chemical complexity of the alloy model.

Finally, we point out a subtle difference in the meaning of $\hat{F}^X$. 
In a pure system (i.e. a single bulk state), $\hat{F}^X(\Delta E_\text{seg}^X)$ is proportional to the number of GB states that correspond to energy $\Delta E_\text{seg}$ for the solute $X$. 
In a solid solution, where $\Delta E_\text{seg}$ consists of bulk and GB energies, $\hat{F}^X$ is proportional to the number of GB and bulk state pairs that yield segregation energy of $\Delta E_\text{seg}$.

\subsection{Segregation energy distribution in thermodynamic models}
\label{sec:McLean_isotherms}
The McLean isotherm~\cite{McLean1957} relates the equilibrium solute concentration for a species $X$ at the GB\footnote{We recall that our model is formulated under the assumption of constant matrix/GB composition and does not treat any co-segregation phenomena.}, $X_\text{GB}^X$, with its bulk concentration, $X_\text{B}$ and the corresponding Gibbs free-energy of segregation, $\Delta G_\text{seg}^X$:
\begin{equation}
    \dfrac{X_\text{GB}}{1-X^\text{GB}} = \dfrac{X^\text{B}}{1-X^\text{B}} \exp\left(-\dfrac{\Delta G^X_\text{seg}}{k_B T}\right) \, .
    \label{eq:mclean}
\end{equation}
In 0\,K first-principles calculation corresponding to ambient conditions, it is common to substitute $\Delta E_\text{seg}^X$ for $\Delta G_\text{seg}^X$ due to the relation
\begin{equation}
    \Delta G = \Delta H -T \Delta S = \overbrace{\Delta U}^{\Delta E} + \overbrace{\Delta p V + p \Delta V}^{0} + \overbrace{T\Delta S}^{0\,\text{K} \Rightarrow 0} \  .
    \label{eq:seg-dis-energy-approximation}
\end{equation}
Note that $\Delta S$ only includes vibrational entropy as the configurational part is already included in McLean equation.  Due to its computational complexity, the vibrational term is usually neglected and $\Delta G_{\text{seg}}^X\approx \Delta E_{\text{seg}}^X$.
For example, for the substitutional segregants in ferrite, this approximation was shown to be reasonable~\cite{Lejcek2021-hl}. In general, vibrational entropy can be expected to cause a further splitting of segregation energies and to reduce segregation trends as discussed recently, e.g., in \cite{Tuchinda2023}.  We do not discuss these effects here but leave them to future investigations. 

To account for the spectral nature of segregation energies, we compute an ``\textit{averaged}'' isotherm by a convolution of the McLean isotherm (Eq.~\eqref{eq:mclean}) with the distribution from Eq.~\eqref{eq:dis-skew-normal}:
\begin{equation}
    \left<\hat{X}^\text{GB}(T)\right> = \int_{-\infty}^\infty X^\text{GB}(\Delta E^X_\text{seg}, T) \hat{F}^X(\Delta E^X_\text{seg}) \text{d} \Delta E^X_\text{seg} \, .
    \label{eq:mclean-averaged}
\end{equation}
We note that the same effective isotherm has been used for also the polycrystalline spectral models~\cite{Wagih2019, Wagih2024-gx, White1977, Muetschele1987}.
Both cases, i.e., polycrystals and chemically complex alloys yield a spectrum of segregation energies related to variety in the sites at the GBs.
The origin of this variety is, however, different: it is of structural character in the polycrystal case and of chemical character in the present study.

In the present work, we compare three different levels of approximation for each solute.
Firstly, we compute the isotherm for a pure Ni system (i.e., using $\Delta E_{\text{seg}}$ determined in a pure Ni matrix). 
In a second step we replace the segregation energy distribution of a real alloy with its mean value (Eq.~\eqref{eq:dis-expectation-contious}) and then use it in single-value McLean isoterm, Eq.~\eqref{eq:mclean}. 
Finally, we compare both with the effective isotherm calculated using Eq.~\eqref{eq:mclean-averaged}.

\subsubsection{Determining the enthalpy and entropy of segregation}
The original purpose of the McLean isotherm was to determine the segregation energy from a set of measured concentrations at different temperatures~\cite{Lejcek2010}. 
By rearranging Eq.~\eqref{eq:mclean} we obtain:
\begin{equation}
    \Delta G^{\text{eff}}_{\text{seg}} = -k_B T\ln{\left(\dfrac{\hat X^\text{GB}(1 - \hat X^\text{B})}{X^\text{B}(1-X^\text{GB})}\right)}\ .
    \label{eq:seg-effective-segregation-energy}
\end{equation}
By substituting Eq.~\eqref{eq:mclean-averaged} for $X_{\text{GB}}$ in Eq.~\eqref{eq:seg-effective-segregation-energy} we obtain a temperature dependence of $\Delta G^{\text{eff}}_{\text{seg}}(T)$. 
We note that the temperature dependence does not relate to an entropy of clear physical meaning but merely originates from averaging the multitude of segregation scenarios due to the distribution of local chemistries (spectrum of segregation energies).
A similar concept has been recently discussed in the literature for the case where the spectrum originated from the geometrical variety of GB structures~\cite{Scheiber2021}.
This is in agreement with the linear temperature dependence of the segregation energy~\cite{Lejcek2010, Erhart1981} often observed in experiments. Therefore, it is the \emph{spectral nature} of segregation energies that gives rise to the temperature dependence $\Delta G^{\text{eff}}_{\text{seg}}$.

In the present case of a multicomponent alloy, Eq.~\eqref{eq:seg-effective-segregation-energy} yields the temperature dependence for a given bulk concentration $X^\text{B}$. 
The slope and intercept of a linear fit will yield an estimate for $\Delta H$ and $\Delta S$ according to Eq.~\ref{eq:seg-dis-energy-approximation}. 
For more details, we refer the reader to ~\ref{sec:app-geff-fit}.

\section{Computational Methods}
\subsection{Atomistic model generation}
The chosen calculation setup involves separate atomistic models for bulk and grain boundary regions.
This is primarily motivated by minimizing the needed computational resources, while maximizing the variety of local environments in the bulk region (i.e. region unaffected by the grain boundary).
We note that many works involving pure metal matrix often employ a single supercell for both regions, e.g., Ref.~~\cite{Razumovskiy2015, Xue2021b}.

\paragraph{Bulk models}
We used a $3\times 3\times 3$ supercell of \textit{fcc}-nickel. 
The resulting 108 lattice sites were populated according to the composition shown in Tab.~\ref{tab:disorder-composition}. 
The atoms were distributed using \textit{sqsgenerator}~\cite{Gehringer2023} with a Monte-Carlo approach, optimizing the short-range parameters on the first seven coordination shells with interaction weights $w^i = \frac{1}{i}$. 
We checked $10^{10}$ configuration to choose a single special quasi-random structure (SQS)~\cite{Wei1990} representing the bulk Ni-based superalloy. 
Subsequently, we placed a solute atom $X$ at each lattice position to sample the bulk states (cf. Sec.~\ref{theory:disorder-at-gb}).

\begin{table}[h]
    \centering
     \begin{tabular}{lcccccc}
        \toprule
         & & Ni & Cr & Co & Ti & Al \\ \midrule
        \multirow{2}{*}{bulk} & $x_{\text{B}}$ & 0.556 & 0.185 & 0.148 & 0.056 & 0.056 \\
        & $N^\text{tot}=108$  & 60 & 20  & 16 & 6 & 6 \\
        \multirow{2}{*}{GB} & $x_{\text{GB}}$ & 0.561 & 0.184 & 0.149 & 0.053 & 0.053 \\
        & $N^\text{tot}=114$ & 64 & 21 & 17 & 6 & 6 \\\midrule
        \multicolumn{2}{c}{$|x_{\text{B}}-x_{\text{GB}}|\ \left[\cdot 10^{-3}\right]$} & 5.85 & 0.975 & 8.29 & 2.92 & 2.92
        \\ \bottomrule
    \end{tabular}
    \caption{Compositions of the bulk and GB simulations cells. 
    For each of them, the upper row is the composition (in at. mole fractions), while the lower gives the number of atoms distributed in the cells. 
    The difference between the bulk and GB cell compositions for each matrix element is given in the last row.}
    \label{tab:disorder-composition}
\end{table}

\paragraph{GB models}
\label{sec:gb-models}
To make our setup comparable with previous literature, we used similar GB cell geometries as reported by~\citet{Razumovskiy2015}. 
The cell vectors $\vec{a}=[1\bar{2}0]$, $\vec{b}=[001]$ and $\vec{c}=[210]$ refer to the axes in Fig.~\ref{fig:model-disordered}. 

We used a vacuum padding of $9.5$~\AA{} in $\vec{c}$ direction. 
In-plane, we created a $2\times3$ ($\vec{a}\times\vec{b})$ supercell (slightly larger than $2\times2$ used in Refs.~\cite{Xue2021b, Razumovskiy2015}). 
Thereby, our GB cells contain 114 atoms.
Each such model contains 30 different GB sites (red region in the left panel of Fig.~\ref{fig:model-disordered}). 
In the pure elemental setting, 12 of them correspond to sites 2 and 3 (marked orange and green in Fig.~\ref{fig:model-disordered}), while 6 are site 1 (blue in Fig.~\ref{fig:model-disordered}). Therefore, we generated five different SQS to sample the GB states in order to sample as many different local environments as possible.
Those were chosen using \textit{sqsgenerator}~\cite{Gehringer2023} by probing $10^{11}$ configurations. 
For the procedure on how we select the five SQS, we refer the reader to \ref{app:selection-sqs}.

We note that the computational complexity for sampling all states of the disordered cell (258 calculations in our particular setup: 108 bulk and 150 GB states) in comparison to a pure metal (one bulk and three GB states) is drastically increased.

Finally, because of the slightly different numbers of atoms in the bulk (108) and GB (114) cells, the composition do not match exactly.
However, the last row of Tab.~\ref{tab:disorder-composition} shows that the maximum deviation is $< 0.5~at.\%$.

\subsection{DFT setup}
The quantum-mechanical calculations were carried out with the Vienna Ab initio Simulation Package (VASP)~\cite{Kresse1993, Kresse1996}. 
For treating exchange and correlation effects, we employed the general gradient approximation as parametrized by Perdew-Burke-Ernzerhof and revised for solids (GGA-PBEsol)~\cite{Perdew2008}.  
For the electronic self-consistent cycle, we set a convergence criterion of $\Delta E_{SCF} =10^{-4}\;\text{eV}$/cell. 
All calculations were carried out in spin-polarized mode.
Spins with an initial magnitude of $2\mu_B$ were ferromagnetically arranged.
The projector augmented wave (PAW) method~\cite{Bloechl1994, Kresse1999} was used to describe the electron-ion interactions. 
For the $k$-mesh sampling of the Brillouin zone, we used a Monkhorst-Pack~\cite{Monkhorst1976} scheme with $4 \times 4 \times 4$ for the bulk, and $4 \times 4 \times 1$ $k$-points for the GB cells. 
We employed a first-order Methfessel-Paxton smearing~\cite{Methfessel1989} with a smearing width of 0.2~eV.
The calculated data are available under the Creative Commons license in the NOMAD archive~\cite{nomad_Ni-based_GB_SQS}.

\section{Results and Discussion}

\subsection{Pure Ni vs. disordered alloy}

\begin{figure}[t]
    \centering
    \includegraphics[width=11cm]{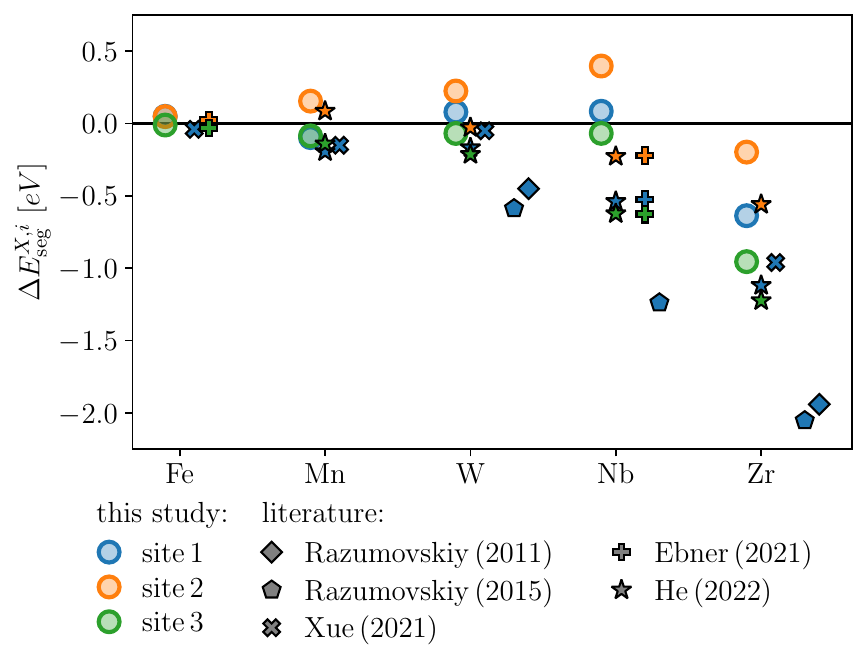}
    \caption{
        Segregation energies to $\Sigma5$ GB in pure Ni. 
        The colors correspond to the sites in Fig.~\ref{fig:model-disordered} (left). The black-outlined symbols represent data from literature (colored by the site type): Razumovskiy (2011)~\cite{Razumovskiy2011b}, Razumovskiy (2015)~\cite{Razumovskiy2015}, Xue (2021)~\cite{Xue2021b}, Ebner (2021)~\cite{Ebner2021}, and He (2022)\cite{He2022-tf}.
    }
    \label{fig:seg-ni-pure}
\end{figure}

Owing to its methodological simplicity, pure elemental Ni is usually taken as a representative model for segregation in Ni-based superalloys. 
Therefore, we also use it as a reference in our study.
Figure~\ref{fig:seg-ni-pure} shows the segregation energy for each solute, Fe, Mn, W, Nb, and Zr, to each GB site in the pure system (Fig.~\ref{fig:model-disordered}, left).
Our data show systematically smaller (more positive) segregation energy than the literature.
We attribute this to the different choice of the XC functional, namely PBEsol in the present study as compared with PBE in all other calculations~\cite{ScheiberPrivate}.
This, for example, leads to a strongly reduced segregation tendency predicted here.
Importantly, the ordering of the site preference, i.e., the most preferable segregation site (with the exception of Mn) is site 3, which is the one further away from the GB plane, is fully consistent with previous reports~\cite{Ebner2021, He2022-tf}.
We therefore conclude that our calculations qualitatively agree with the previous reports, and we can proceed in discussing the impact of real alloy composition on segregation.

\begin{figure}[t]
    \centering
    \includegraphics[width=\textwidth]{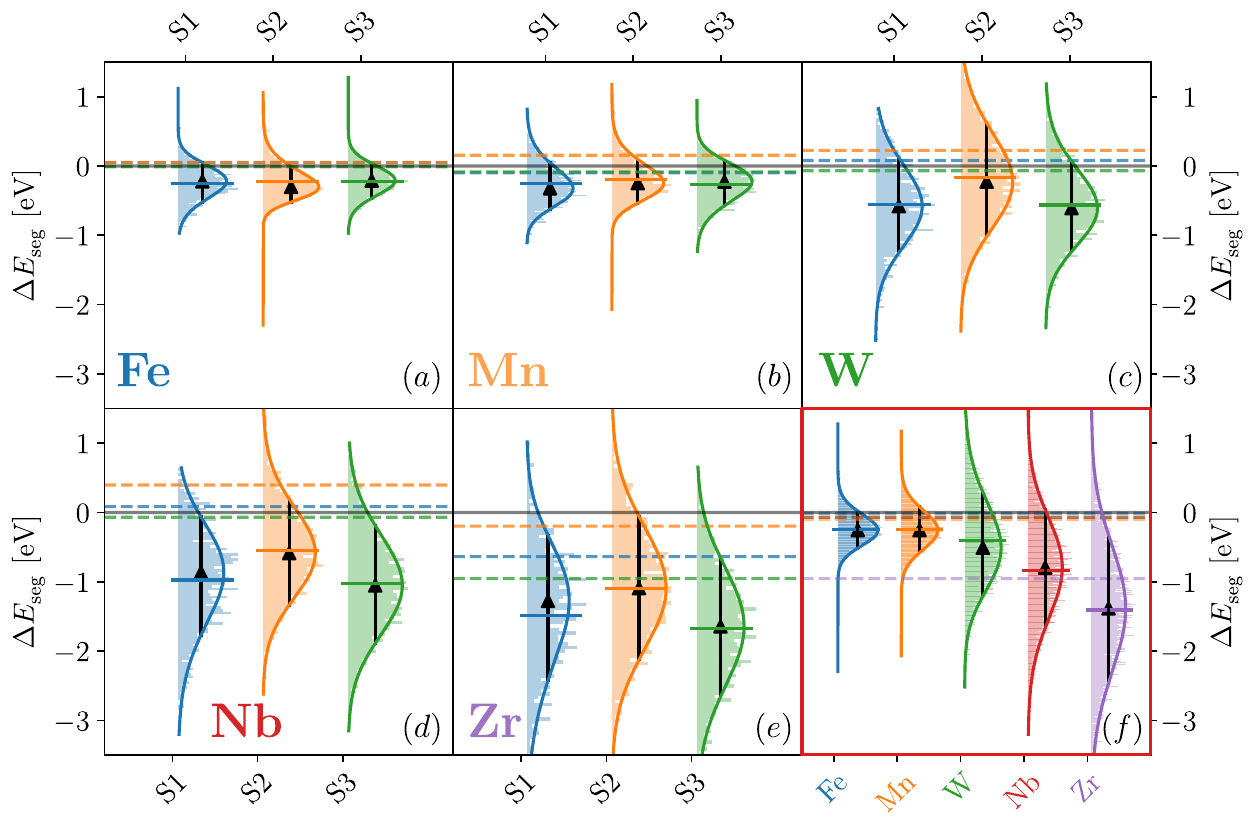}
    \caption{
        Panels (a)--(e) show the segregation energy spectra for each site per solute.
        The colors of the spectra refer to Fig.~\ref{fig:model-disordered} (right). 
        The solid colored horizontal line is a mean value of the corresponding distribution while the black line and symbol visualize the $\sigma$ and $\mu$ parameters of the skew-normal fit (Eq.~\eqref{eq:dis-skew-normal}, shown with the colored curve) to a discrete histogram. 
        The dashed horizontal lines are the energies for the pure GB (see Fig.~\ref{fig:seg-ni-pure}).         
        The last panel (f) shows the merged distributions for each solute; e.g., the distribution for Fe in (f) is obtained by merging the three distributions S1--S3 from (a). 
        The color code in panel (f) is consistent with Fig.~\ref{fig:formation-energies-splitted}.
    }
    \label{fig:seg-ni-disordered}
\end{figure}

We now turn our attention to a model of the real disordered alloy.
Figure \ref{fig:seg-ni-disordered} shows the spectra of segregation energies (Eq.~\eqref{eq:segregation-energy-definition}) together with the skew-normal fits for all five solutes. 
Each of the plots shows three spectra, one for each of the three types of GB sites as defined in pure Ni (S1, S2, and S3). 
The dashed horizontal lines in Fig.~\ref{fig:seg-ni-disordered} represent the pure Ni reference values (cf. Fig.~\ref{fig:seg-ni-pure}).
All values (mean, fit parameters, and pure segregation energies) for Fig.~\ref{fig:seg-ni-disordered} are summarized in the Appendix, Table~\ref{tab:dis-fit-parameters-site}.

Figs.~\ref{fig:seg-ni-disordered}a--\ref{fig:seg-ni-disordered}e reveal three major insights. 
Firstly, a comparison of the expectation values $\left<\Delta E^X_\text{seg}\right>$ (solid colored horizontal line) of the disordered alloy with the corresponding segregation energies in pure Ni case (colored dashed lines) yields a drastically enhanced (more negative) tendency in the former. 
While for iron, this enhancement is $\approx 0.25$~eV for all three segregation spectra (cf.~Fig.~\ref{fig:seg-ni-disordered}a), we find up to $\approx 1$~eV for Nb (Fig.~\ref{fig:seg-ni-disordered}d). 
Despite this enhancement, the qualitative behavior of individual sites is preserved.
This is particularly obvious in the cases of W, Nb, and Zr, where the mean value corresponding to the S3 sites is clearly lower than that of the S1 and S2 sites, where the latter is the least favored scenario.

In order to explain the enhanced segregation for the complex matrix compared to pure Ni we consider phenomenological segregation models. The segregation energy is determined chiefly by two terms, a bonding contribution and an elastic contribution. The former arises mainly from the change in cohesive energy between solute and matrix and to a weaker extent from interaction of solute and matrix. The elastic contribution arises from the volume difference between solute and matrix. The two contributions seem to cancel almost exactly for W in pure Ni. The higher cohesive energy of W causes anti-segregation (positive segregation energy) while the higher atomic radius would cause segregation (negative segregation energies). The Ni-Cr-Co-Ti-Al solid solution has a higher volume ($V = 10.737\,$\AA{}$^3$ vs. $V=10.253\,$ \AA{}$^3$ for pure Ni) and therefore, the elastic contribution is reduced. Since, overall, the opposite is observed we conclude that the bonding contribution mainly causes the enhanced segregation. This would imply that the cohesive energy of the complex matrix is higher than the one of Ni, and the bonds formed between Ni and W are stronger than the bonds between the complex matrix and W. Similar considerations should apply for the other solutes.

Secondly, for Fe, Mn, and Zr, we find nearly the same FWHM irrespective of the segregation site S1, S2, or S3 (all three values are within 0.1~eV for each species). 
W exhibits $\approx 0.3$~eV broader distribution for S1 as compared with those of S2 and S3; contrarily, the S2 spectrum of Nb is $\approx 0.2$~eV narrower than the S1 and S3 spectra. 
Thirdly, the fitted skew values $\alpha$ listed in Tab.~\ref{tab:dis-fit-parameters-site} clearly reveal that a Gaussian distribution (as used in Ref.~\cite{Huber2018}) is not sufficient to describe any of the spectra.
We recall that according to Eq.~\eqref{eq:dis-skew-normal}, the skew-normal distribution becomes a Gaussian distribution for $\alpha=0$, while all our fits yield $|\alpha|>0$. 
Similarly, also a Gumbel distribution (as used in Ref.~\cite{Scheiber2021}) does not have enough degrees of freedom. 
This is demonstrated, e.g., by a sign change of the skew values for Fe of S1 ($\alpha < 0$) and S2 ($\alpha > 0$). 
In other words, the skewness as a degree of freedom is needed to describe the left-skewed S1 and right-skewed S2 spectrum. 

\begin{table}[b]
    \centering
    \begin{tabular}{cccccc}
        \toprule
         & Fe & Mn & W & Nb & Zr \\ \midrule
         $\min\Delta E_\text{seg}^\text{pure}$~[eV] & $-0.01$ & $-0.10$ & $-0.07$ & $-0.07$ & $-0.95$ \\
         $\left<\Delta E_\text{seg}^X \right>$~[eV] & $-0.24$ & $-0.24$ & $-0.42$ & $-0.83$ & $-1.41$ \\
         $\alpha$~[-] & 0.97 & 0.72 & 1.45 & $-0.94$ & $-0.66$ \\
         FWHM~[eV] & 0.53 & 0.66 & 1.49 & 1.70 & 2.14 \\ \bottomrule
    \end{tabular}
    \caption{
        Fitting parameters of the merged segregation energy spectra (combination of S1, S2, and S3 as presented in Fig.~\ref{fig:seg-ni-disordered}f). 
        The first row shows the minimum segregation energy values per solute for pure Ni (Fig.~\ref{fig:seg-ni-pure}).
    }
    \label{tab:dis-fit-parameters}
\end{table}

However, the differentiation between S1, S2, and S3 has been made solely for better comparison with pure Ni. 
In a disordered system, every site in the GB zone (red region in Fig.~\ref{fig:model-disordered}) is generally surrounded by different matrix species, leading to largely overlapping spectra for the sites S1--S3 (Fig.~\ref{fig:seg-ni-disordered}a--e).
Consequently, we consider only a single spectrum for each solute. 
Those are shown in Fig.~\ref{fig:seg-ni-disordered}f, where each spectrum is computed by merging the three spectra in the corresponding panel.
For example, the blue spectrum for Fe in Fig.~\ref{fig:seg-ni-disordered}b is obtained by merging the three spectra from Fig.~\ref{fig:seg-ni-disordered}a.
Again, we fitted the resulting spectra with skew-normal distributions (shown in Fig.~\ref{fig:seg-ni-disordered}f) and the resulting fitting parameters present in Tab.~\ref{tab:dis-fit-parameters}.
Comparison with the (lowest) segregation energies in pure Ni (first row) confirms the significant enhancement due to the chemical disorder.
This enhancement is up to an order of magnitude for Fe, W, and Nb.
For example, while Nb exhibits nearly no tendency to segregate to the GB in pure Ni ($\Delta E_\text{set}^\text{pure}= -0.07~\text{eV}$), a mean value of the alloy segregation spectrum is $\left<\Delta E_\text{seg}^\text{Nb}\right> = -0.83~\text{eV}$. 
In contrast, for Zr we report $\Delta E_\text{seg}^\text{pure}= -0.95~\text{eV}$ and hence already a strong segregation tendency in pure Ni, but we still predict an enhancement to $\left<\Delta E_\text{seg}^\text{Zr}\right> = -1.41~\text{eV}$ for the alloyed system.

Finally, we note that the spectral properties of the segregation energy cannot be ignored. 
The distributions shown in Fig.~\ref{fig:seg-ni-disordered}f are too broad to be replaced with a mean value. 
In particular, the FWHM of all the spectra is in the range or larger than its mean value. 
For example, for W, we obtain a mean segregation energy of $\left<\Delta E_\text{seg}^\text{W}\right> = -0.42~\text{eV}$, whereas its FWHM is $1.49~\text{eV}$. 
Furthermore, our (limited) data do not suggest any trend between the mean values and the FWHM. 
For example, the mean value $\left<\Delta E_\text{seg}\right>$ is nearly twice as low for Nb compared to W, the FWHM of Nb increases only slightly w.r.t. W (cf. Tab.~\ref{tab:dis-fit-parameters}).

\subsection{Thermodynamics of segregation}
\label{sec:results-impact-mclean}

\begin{figure}[ht!]
    \centering
    \includegraphics[width=\textwidth]{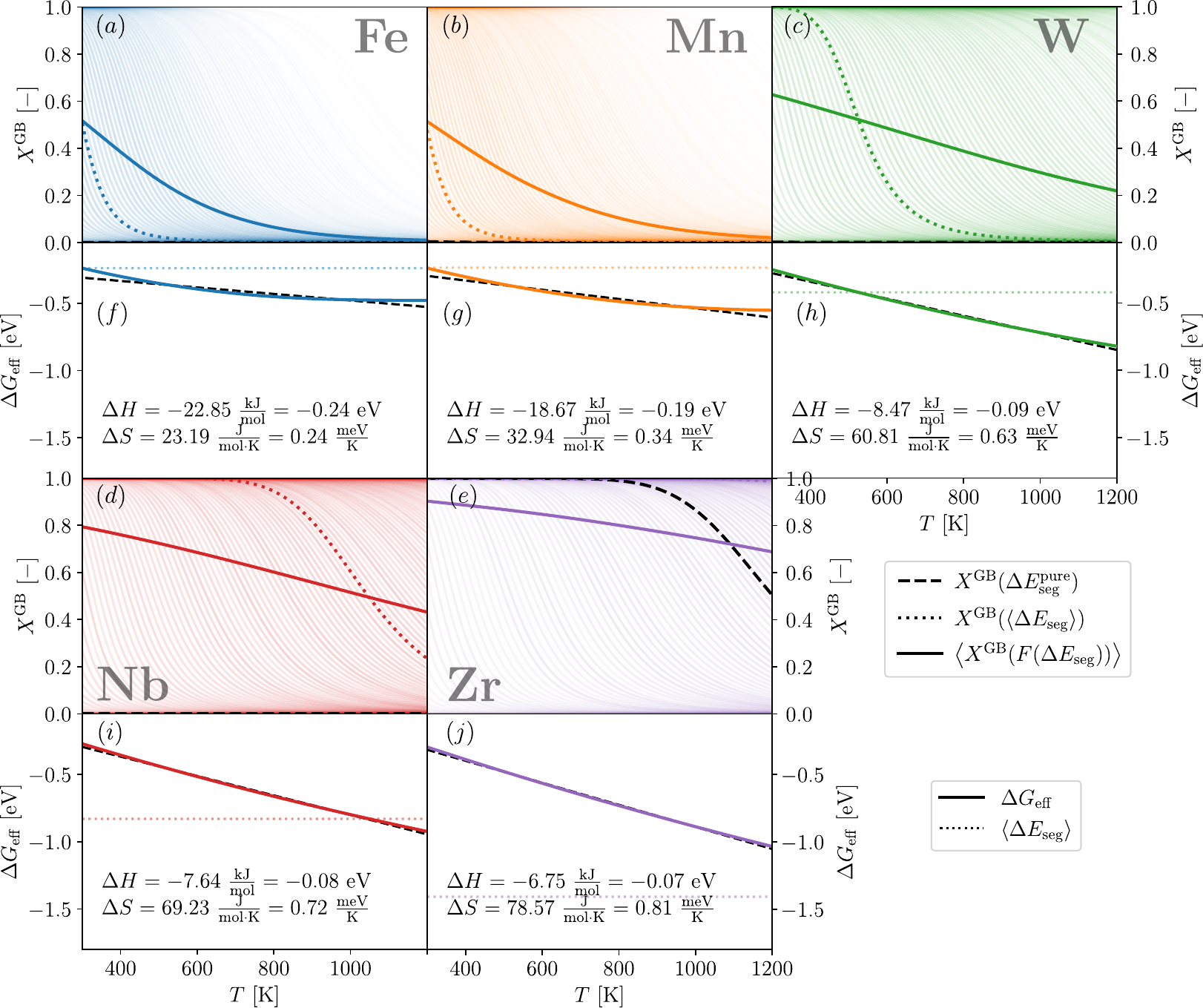}
    \caption{The upper panels (a)--(e) show the impact of the energy spectra on McLean-type segregation. 
    The black dashed (often overlapping with the $x$-axis) and colored dotted lines show the isotherms calculated according to Eq.~\eqref{eq:mclean} for the pure system and the mean value of the distribution (Eq.~\eqref{eq:dis-expectation-contious}), respectively. 
    The translucent isotherms in the background refer to the different energy states (segregation scenarios) of the histogram. 
    The colored solid line is the average of those isotherms according to Eq.~\eqref{eq:mclean-averaged}. 
    The solid lines in the lower panels (f)--(j) are the effective temperature-dependent free energies of segregation, $\Delta G^\text{eff}_{\text{seg}}(T)$ (Eq.~\eqref{eq:seg-effective-segregation-energy}). 
    The black dashed lines are linear fits between 0 and 1200\,K, with the corresponding enthalpy $\Delta H$ and entropy $\Delta S$ values given. 
    The colored dotted horizontal line is $\left<\Delta E^X_\text{seg}\right>$ (Eq.~\eqref{eq:dis-expectation-contious}) as a reference. 
    All isotherms in panel (a)--(e) are plotted for constant $X^\text{B} = 0.01~\text{at.}~\%$, i.e. 100 ppm.}
    \label{fig:mclean-isotherms}
\end{figure}

The segregation energetics discussed in the previous section serve as inputs to the thermodynamic assessment of grain boundary segregation using McLean isotherms described in Sec.~\ref{sec:McLean_isotherms}.
These predict the fraction of GB sites a segregating species occupies at a given temperature.
The results are summarized in Fig.~\ref{fig:mclean-isotherms} for an example of the dilute limit with a bulk concentration of $X^B=0.0001\,\mathrm{at.\%}=100\,\mathrm{ppm}$.
For each species, the black dashed line is McLean isotherm corresponding to the minimum segregation energy in pure Ni (Fig.~\ref{fig:seg-ni-pure} and Tab.~\ref{tab:dis-fit-parameters}).
The McLean isotherms based on mean values of the segregation spectra (Eq.~\eqref{eq:mclean-averaged}) of Ni-based disordered alloy are shown with colored dotted lines.
We recall that those values are significantly lower (i.e., representing stronger segregation tendency) than for the case of pure Ni (cf.~Tab.~\ref{tab:dis-fit-parameters}).
Consequently, significantly higher solute concentrations in the GB sites are predicted for the disordered alloy compared with the pure Ni case, and the GB sites retain their full occupancy by the solutes ($X^{GB}\approx1$) to higher temperatures.

In contrast, the single isotherm computed from the mean value spectrum (colored dotted line) overestimates the GB concentration at lower temperatures but drops below the averaged isotherms, as those show a significantly flatter slope. 
The flattening of the averaged isotherms becomes more pronounced for solutes with an increased segregation tendency (e.g., compare Fe and Mn on the one hand, with W, Nb, and Zr on the other hand) while the crossover between the mean and the averaged isotherm shifts to higher temperature. 
Consequently, this crossover is not in the shown temperature range for Nb and Zr anymore. 
In summary, all three isotherms are significantly distinct from each other, e.g., for W we find at 700~K concentrations ranging from $X^\text{GB}(\Delta E_\text{seg}^\text{pure}) = 0.0311\,\% \ 311\,\text{ppm}$ over $X^\text{XB}(\left<\Delta (E_\text{seg})\right>) = 9.7\,\%$ to $\left<X^\text{GB}(F(\Delta E_\text{seg})) \right> = 43.6\,\%$. 
It is useful to remind us what these numbers mean: they provide a fraction of GB sites (arbitrarily assigned in our model to be the three closest planes to the GB, cf. Fig.~\ref{fig:model-disordered}) occupied by the minor species $X$.
It becomes obvious that these concentrations are not dilute anymore and that a natural extension of this model would be to include (at least) solute-solute interactions.
Nevertheless, these results also clearly demonstrate the hugely different behavior related purely to the chemical disorder of the matrix.
Finally, we note that while the cases corresponding to a single-valued segregation scenario (pure Ni and $\langle \Delta(E_{\text{seg}})\rangle$) lead to a 100\,\% GB site occupancy at $T\to0\,\text{K}$, in the case of spectral segregation energies containing also a fraction of anti-segregating sites (cf. Fig.~\ref{fig:seg-ni-disordered}), the fraction of occupied GB sites is smaller than 1 even at 0\,K.

The lower panels show the effective segregation energy (solid color lines) according to Eq.~\eqref{eq:seg-effective-segregation-energy}. 
For each graph, the dotted line gives a mean value of the distribution as a reference. 
The black dashed line is a linear fit of $\Delta G^{\text{eff}}_{\text{seg}}(T)$ to the temperature range between 0 and 1200\,K.
This linear fit allows extraction of enthalpy, $\Delta H$, and entropy, $\Delta S$, of segregation. 

The reports on experimentally measured data of segregation enthalpies and/or entropies for pure Ni or Ni-based alloys are scarce.
Muschik et al.~\cite{Muschik1989-en} reported $\Delta H\approx (38\pm3)\,\text{kJ/mol}$ and $\Delta S=(-0.5\pm0.5)R$ for symmetric grain boundaries formed as bi-crystals of Ni-In.
A value of $(50\pm5)\,\text{kJ/mol}$ was reported for Ni-In polycrystals at 970\,K.
From this comparison, the enthalpy and Gibbs free energy of segregation are in the same order of magnitude, while our values of entropy of segregation are an order of magnitude larger.
A comprehensive overview of these quantities has been collected for Fe-based systems by Lej\v{c}ek and co-workers~\cite{Lejcek1991, Lejcek2001, Lejcek2016}.
Our DFT-based predictions are in the same order of magnitude as those found for substitutional solutes in $\alpha$-iron.

Although the linear fits seem relatively representative in the shown regime, they should be taken with a grain of salt.
Firstly, the fit is not ideal (as evidenced for example for Fe, Fig.~\ref{fig:mclean-isotherms}f).
This suggests that $\Delta H$ and $\Delta S$ are actually not constants but rather should be treated as $T$-dependent.
Secondly, the actual values of $\Delta H$ and $\Delta S$ strongly depend on the chosen fitting range (here 0--1200\,K).
This is a consequence of the non-linearity of the $\Delta G_{\text{eff}}(T)$ curve.
Finally, also the overall/bulk concentration $X^B$ influences the extracted $\Delta G_{\text{eff}}$, and hence $\Delta H$ and $\Delta S$ also become concentration dependent.
These facts make it difficult to interpret them in the framework of classical thermodynamics with single-valued entropy and enthalpy segregation values or to compare them to Table~\ref{tab:dis-fit-parameters}.

\section{Conclusions}
In the present article, we described a methodology for modeling segregation in a multi-component disordered solid solution. 
We proposed a novel approach that is based on well-established models and allows to calculate the segregation energy distributions for solutes in the dilute limit in compositionally complex systems.

We applied this method to the segregation of Fe, Mn, Nb, W, and Zr in a model system inspired by Ni-based superalloys. 
Importantly, we showed that first-principles predictions for disordered models lead to qualitatively different results than for pure Ni. 
To quantify the differences, we extensively discussed the segregation energy spectra, thereby highlighting their essential importance.

In the second part, we compared the impact on the predictions based on the McLean model. 
We showed that even when replacing the distribution with a single value---the mean of the distribution---we predicted qualitatively different behavior compared to pure Ni.
Next, we presented a complete spectrum of isotherms based on the McLean model, corresponding to the spectrum of the segregation energies.

We reiterate that the here-reported segregation enthalpy and entropy are consequences of the chemical complexity of the matrix material; further level of complexity would stem from the geometrical variety of grain boundary structures.
The present model deals with the segregation of minor species (with dilute concentrations) and fixes the composition of the matrix.
A further step towards describing segregation phenomena in real compositionally complex alloys is the segregation of matrix elements.
Since these are concentrated, approaches beyond the dilute limit are required, as recently shown, e.g., in Ref.~\cite{Spitaler2025-vb}.
In fact, these should be included even in the dilute limit for low temperatures, where the GB site occupancy strongly increases.
Another topic specifically important for chemically complex materials, includes site-competition phenomena.
Nevertheless, the present work clearly demonstrates the importance of the spectrality of the segregation energies related to the chemical disorder.

\section*{Acknowledgements}
D.G. greatly appreciates the support (DOC scholarship) from the Austrian Academy of Sciences (\"OAW). D.G. and D.H. acknowledge financial support by \"Oster\-rei\-chi\-sche For\-schungs\-f\"or\-der\-ungs\-ge\-sell\-schaft mbH (FFG), project number FO999888151, ``AMnonWeldSuperAlloys''.

The computational results were in part achieved by using the VSC computing infrastructure. 
The authors also sincerely thank  V.~I. Razumovskiy and D. Scheiber from the Materials Center Leoben (MCL) Forschung GmbH, and S. Bodner from Montanuniversität Leoben for their input and helpful discussions.

\section*{Declaration of Competing Interest}
The authors declare that they have no known competing financial interests or personal relationships that could have appeared to influence the work reported in this paper. 

\appendix

\section{Selection process for the five SQS}
\label{app:selection-sqs}
The SQS optimization routine~\cite{Gehringer2023} will result in multiple candidate structures.
We want to minimize the number GB states that are coordinated similarly and so to maximize the range of different local environments.
Therefore, let $i \in \{1,\ldots,30\}$ be a GB site in a candidate SQS GB structure. 
Then, we represent the local environment $x_i$ by a histogram of the neighboring species
\begin{equation}
    x_i = \{N^{\widetilde M}_i \}_{\forall \widetilde M \in \mathbf{M}}\ ,
\end{equation}
where $N^{\widetilde M}_i$ denotes the number of $\widetilde M$ atoms in the first coordination shell of the $i^\text{th}$ site. 
The set of all local environments $X_\xi$ is then
\begin{equation}
    X_\xi = \{x_i\}_{\forall i = 1,\ldots, 30}  \, ,
\end{equation}
where $\xi$ is the index of a site in the SQS structure. 
$i$ can take 30 values since for our particular setup (Sec.~\ref{sec:gb-models}), the GB (red) region in Fig.~\ref{fig:model-disordered} contains that many sites.
Hence, we want to find several (in our case 5, $\xi=1,\dots,5$) different SQSs, that maximize
\begin{equation}
    \left| \bigcup_\xi X_\xi \right| \to \max\, .
\end{equation}
By sampling 150 ($5 \times 30$) GB states, we could identify 5 SQS cells that yield together 133 differently coordinated sites in the first coordination shell.
The local chemical compositions, re-calculated to only the GB zone, are summarized in Tab.~\ref{tab:gb-zones}.

\begin{table}[h]
    \centering
    \begin{tabular}{lcccccc}
        \toprule
          & Ni & Cr & Co & Ti & Al & Note \\ \midrule
        nominal & 55.56 & 18.52 & 14.80 & 5.56 & 5.56 & see Tab.~\ref{tab:nominal-composition} \\ \midrule
        SQS 1 & 66.67 & 10.0 & 10.0 & 0.0 & 13.33 & Ni$\uparrow$, Cr$\downarrow$, Co$\downarrow$, Ti$\downarrow\downarrow$, Al$\uparrow\uparrow$ \\
        SQS 2 & 60.0 & 6.67 & 16.67 & 16.67 & 0.0 & Ni$\uparrow$, Cr$\downarrow$, Cr$\uparrow$, Ti$\uparrow\uparrow$, Al$\downarrow\downarrow$ \\
        SQS 3 & 60.0 & 20.0 & 10.0 & 6.67 & 3.33 & Ni$\uparrow$, Cr$\approx$, Co$\downarrow$, Ti$\approx$, Al$\approx$ \\
        SQS 4 & 53.33 & 26.67 & 10.0 & 0.0 & 10.0 & Ni$\approx$, Cr$\uparrow$, Co$\downarrow$, Ti$\downarrow\downarrow$, Al$\uparrow$ \\
        SQS 5 & 36.67 & 26.67 & 23.33 & 0.0 & 13.33 & Ni$\downarrow\downarrow$, Cr$\uparrow$, Co$\uparrow$, Ti$\downarrow\downarrow$, Al$\uparrow\uparrow$ \\\bottomrule
    \end{tabular}
    \caption{Compositions GB zones (red region in Fig.~\ref{fig:model-disordered}) for the five different SQSs. The first row shows the overall composition of the GB (see Tab.~\ref{tab:nominal-composition}). Note that the GB zone contains only 30 sites, therefore, the ``\textit{compositional accuracy}'' is limited to $\approx 3.33$~at.\%.}
    \label{tab:gb-zones}
\end{table}

\clearpage

\section{Fitting parameters for Fig.~\ref{fig:seg-ni-disordered}}
\label{app:pure-ni}

\begin{table}[ht!]
\centering
    \begin{tabular}{@{}ccccccc@{}}
        \toprule
         & Site & Fe & Mn & W & Nb & Zr \\ \midrule
         \multirow{3}{*}{$\Delta E_\text{seg}^\text{pure}$~[eV]} & S1 & 0.05 & $-0.10$ & 0.08 & 0.08 & $-0.64$ \\
         & S2 & 0.05 & 0.15 & 0.22 & 0.40 & $-0.20$ \\ 
         & S3 & $-0.01$ & $-0.09$ & $-0.07$ & $-0.07$ & $-0.95$ \\\cmidrule(l){2-7} 
         \multirow{3}{*}{$\left<\Delta E_\text{seg}^X \right>$~[eV]} & S1 & $-0.26$ & $-0.28$ & $-0.60$ & $-1.00$ & $-1.46$ \\
         & S2 & $-0.24$ & $-0.19$ & $-0.19$ & $-0.56$ & $-1.11$ \\ 
         & S3 & $-0.23$ & $-0.28$ & $-0.59$ & $-1.05$ & $-1.70$ \\\cmidrule(l){2-7}
         \multirow{3}{*}{$\alpha$~[-]} & S1 & $-1.40$ & 1.77 & 0.81 & $-1.43$ & $-1.81$ \\
         & S2 & 2.30 & 1.66 & 1.05 & 0.87 & $-0.01$ \\ 
        & S3 & $-0.86$ & $-1.44$ & 1.06 & 0.79 & $-0.78$ \\\cmidrule(l){2-7} 
         \multirow{3}{*}{FWHM~[eV]}& S1 & 0.57 & 0.67 & 1.37 & 1.71 & 2.10 \\
         & S2 & 0.52 & 0.63 & 1.64 & 1.53 & 2.06 \\  
         & S3 & 0.50 & 0.66 & 1.30 & 1.70 & 2.00 \\\bottomrule
    \end{tabular}
    \caption{Expectation values of the energy spectra $\left<\Delta E_\text{seg}^X \right>$~(Eq.~\eqref{eq:dis-expectation-contious}) and skew parameter $\alpha$ and FWHM of the fitted skew-normal distribution for each type of segregation site separately. 
    The first three rows show the segregation energy in the pure Ni system $\Delta E_\text{seg}^\text{pure}$, for each of the pristine GB sites. 
    The columns correspond to Fig.~\ref{fig:seg-ni-disordered}a--e.}
    \label{tab:dis-fit-parameters-site}
\end{table}

\clearpage

\bibliographystyle{elsarticle-num-names} 
\bibliography{cas-refs}

\end{document}